\documentstyle[11pt,newpasp,twoside,epsf]{article}
\def\approxgt      {\lower.2em\hbox{$\buildrel > \over \sim$}}
\def\approxlt      {\lower.2em\hbox{$\buildrel < \over \sim$}}

\def \ms           {\hbox{M$_{\odot}$}}           
\def \kms       {km~s$^{-1}~$}

\def\edcomment#1{\iffalse\marginpar{\raggedright\sl#1\/}\else\relax\fi}
\reversemarginpar

\begin{document}
\title{Molecular Gas Concentrations Outside the Merging Disks}
 \author{Yu Gao$^1$, J.D. Goldader$^2$, E.R. Seaquist$^3$, \& Cong Xu$^1$}
\affil{1 IPAC, Caltech, MS 100-22, 770 S. Wilson Ave., Pasadena, CA 91125}
\affil{2 University of Pennsylvania, Department of Physics and Astronomy,
209 S. 33rd St., Philadelphia PA 19104}
\affil{3 University of Toronto, Department of Astronomy,
60 St. George St., Toronto, ON M5S 3H8, CANADA}

\begin{abstract}
We present our preliminary BIMA CO(1-0) images of II~Zw~96
which show huge molecular gas concentrations outside the 
merging disks. The dominant extra-disk CO 
concentrations in II~Zw~96 correspond to the two star-forming ``knots''
hidden by dust, whereas the others are associated with the two spiral 
disks. The unique huge CO concentrations, $\approxgt 5$ kpc
away from the merging nuclei, account for
$\approxgt 50 \%$ of the total CO emission, yet only at 
a moderate gas surface density as compared with that of ultraluminous
infrared galaxies (ULIGs). 
We tentatively conclude that the huge extra-disk gas concentrations 
are most likely the overlap region of the gas disks,
not necessarily coinciding with the stellar disks overlap region. 
The gas disks overlap region could be 
formed prior to the merging of the stellar disks and 
become the overlap regions found between the disks like
that in Arp~244 (the ``Antennae'' galaxies). These
overlap regions would eventually nurture the {\it extreme starbursts} 
and evolve into the extended gas disk surrounding the double nuclei
in final merging in most ULIGs.
\end{abstract}

\section{Heavily Obscured Extra-disk Star-Forming Regions}

Starburst galaxy mergers involve highly dissipative collapse of
tremendous amounts of molecular gas from their location in two separate
galaxies, into the inner few 100 pc of the  
merged system. The high gas densities in the core of the merger prompt
powerful star formation; the dust associated with the gas absorbs
the UV/optical light and re-radiates in the far-IR, resulting in 
an ultraluminous infrared galaxy (ULIG).
 
In most ULIGs, merging is sufficiently advanced that the gas has 
mostly collapsed into the central a few 100 pc. Notable examples
are the nearest ULIGs like Arp~220 (Downes \& Solomon 1998 (DS98);
Sakamoto et al. 1999). There are, however, a few very highly 
luminous galaxies with extranuclear
star forming regions (just below the $10^{12}L_{\odot}$ ULIG 
cutoff) where the gas has apparently not yet collapsed into 
the nuclei (e.g., Arp~299 Aalto et al. 1997;
Casoli et al. 1999; VV~114 Yun et al. 1994).
These systems are of crucial importantance as they give us a glimpse
into the (apparently brief) intermediate stage between well-separated
gas-rich pairs and the merged late-stage ULIGs.

II~Zw~96 ($L_{\rm IR}=7.6\times 10^{11}L_{\odot}$, nuclear separation
$\sim 7.5$ kpc), shows four distinct star-forming regions, as 
revealed by optical and near-IR
observations (Goldader et al. 1997 (G97)). Two of those are
apparently associated with galaxy nuclei.  The other two sources
lie well away from the two nuclei, extremely weak in optical,
but prominent in near-IR, implying heavy dust obscuration in 
these extra-disk star forming regions. 

\section{Huge Extra-Disk Molecular Gas Concentrations}

In order to explore the extra-disk star forming regions
heavily obscured by dust, we have observed II~Zw~96 at CO(1-0) 
using the 10-element BIMA array at Hat Creek, California.
The data presented here are mainly compact 
C and D array observations plus some limited 
long baseline B array observations. The combined data 
give a synthesized beam of $5.3'' \times 4.2''$ ($3.5\times 2.8$ kpc).
Additional B array observations at $\sim 2''$ resolution are being 
conducted currently to further improve the resolution. 

\begin{figure}
\plotfiddle{gaoy1.eps}{2.25in}{-90}{45}{45}{-160}{240}
\caption{Integrated CO intensity contours from BIMA D, C \& B array 
observations, overlaid on grey scale
H-band image of II~Zw~96. Contours start at $2\sigma$ and increase
by 1$\sigma=1.4$ Jy/beam~\kms}
\end{figure}

A total H$_2$ mass of 
$\sim 2.8 \times 10^{10} M_{\odot}$ (standard conversion) 
was estimated from a single-dish CO survey (Gao~\& Solomon 1999). 
Our BIMA observations have recovered most of the single-dish flux 
and revealed that II~Zw~96 is a morphologically 
complex and extraordinary system with most of CO emission 
located outside the two merging stellar disks (Fig.~1). 
The dominant CO concentrations correspond to the two extra-disk
star-forming regions hidden by heavy dust, 
whereas the others are associated with the two galaxy 
disks and the region between them. 
The huge CO concentrations far away from the galaxy disks
account for more than $\sim 50 \%$ of the total CO emission
with a H$_2$ mass of $\sim 1.4 \times 10^{10} M_{\odot}$.

Huge concentrations of molecular gas in the overlap regions
have been observed already in merger systems like Arp~299
and the ``Antennae'' galaxies Arp~244. The enormous amount of
molecular gas located far away from merging disks makes
II~Zw~96 quite unique. Moreover, the gas density
of $\approxgt 5.3 \times 10^2 \ms/pc^2$ (as seen by a beam resolution
of $3.5 \times 2.8$ kpc) at the peak of the concentrations, 
although at least $50 \%$ higher than either nuclear regions
of the merging galaxies, is orders of magnitude less than
that of ULIGs and even Arp~299. The peak gas surface density
would still be several times less than that of ULIGs even if 
the entire CO emission covered by current larger synthesized
beam is from the central $\approxlt 1$ kpc 
region. In addition, the extra-disk
gas concentrations appear to be peaked with a small offset 
from the star forming knots identified in near-IR (Fig.~1)
and are mostly at higher velocities (Fig.~2). Only a small amount
of molecular gas has been observed in SE galaxy
which appears to have little association with the extra-disk
gas concentrations (Fig.~2). 

\begin{figure}
\plotfiddle{gaoy2a.eps,gaoy2b.eps}{2.25in}{-90}{45}{45}{-160}{240}
\caption{Velocity channel maps at spacing of 20 \kms overlaid on an 
H-band image. The contours are 3, 4, 5.6, 8, 11.3 
($\times \sigma=14$~mJy/beam) and zero velocity corresponds to 10900 \kms.}
\end{figure}

NW galaxy has a rather extended gas distribution, gas rotation motion 
and connection of the disk gas with the extra-disk CO concentration, 
whereas much less gas is observed
in the SE disk. It is difficult to speculate that most of the gas 
in the extra-disk concentrations could have come from the NW galaxy.
Could the 
SE galaxy have been originally very gas-rich with extended 
gas disk and left most of the gas behind after colliding
with the NW galaxy ?

\section{Evolution of the ``Interaction Zones''}

The star forming knots seen in near-IR in the extra-disk gas 
concentrations show properties of the H{\sc ii} regions 
as that of the SE nuclear region indicating powerful young 
starbursts (G97).
These same spectral features are also found in Arp~299 and VV~114.
In all these systems, the apparent ages of the starbursts are $\sim$8--30 Myr,
much younger than the allowed range of ULIGs.
In fact, Goldader et al. (G97) suggested that these three systems
might be examples of a short-lived phase through which 
many---perhaps all--- pass on their way to becoming more
advanced, centrally concentrated ULIG mergers.

Because the HI disks are more extended than the stellar disks, 
they may merge first prior to the merging of the stellar 
and molecular gas disks when a pair of gas-rich spirals approach
each other. Indeed, we may
have seen such early merging and formation of the HI disks overlap
region in NGC~6670 where huge concentration of HI 
has been observed between the two galaxies whose HI disks have
been highly disrupted, whereas little distortion is observed in
the stellar disks (Wang et al. 2000). 
One plausible idea is that the HI gas disk overlap regions formed
much earlier in early mergers (NGC~6670) may be the progenitors
of the apparent overlap regions in intermediate mergers (Arp~244), 
which may evolve into the extended gas disks around the 
much more concentrated double nuclear gas disks in ULIGs (Arp~220, 
DS98). II~Zw~96 fits in between early mergers
like NGC~6670 and intermediate mergers like Arp~244.

As merging progresses and the gas overlap region further concentrates, 
these huge CO concentrations will 
eventually merge with the double nuclear gas disks. They could thus, 
be the progenitors of the so called {\it extreme starburst} regions, 
which dominate the overall starburst
power output of the entire system, yet be confined within sub-kpc
scale (DS98). They may also appear to be
the multiple ``nuclei'' claimed in some ULIGs (cf. Borne et al. 2000).

\section{ALMA Implications}

Although there are now several galaxy mergers where huge
gas concentrations in the overlap regions -- the ``interaction 
zones'' have been observed, II~Zw~96 appears to be the only
one with most of the gas concentrations (presumably the
overlap regions) located outside the stellar merging disks.
Is this unique~? Are there more such systems to be found,
especially in the distant universe where galaxy collisions
are expected to be much more frequent~?
Perhaps, one should not be totally surprised 
when ALMA has mapped some huge molecular gas concentrations with 
little or no optical correspondences, or the peaks of the molecular 
gas concentrations offset from the optical peaks by several kpc,
in merging/star-forming galaxies at high redshift.

\end{document}